\documentclass[prd,aps,a4paper,nofootinbib,showpacs,showkeys,twocolumn]{revtex4}  

\newif\ifusesec
\usesectrue  
   
\usepackage{graphicx} 
\usepackage{mathrsfs}
\usepackage{amsmath,amsfonts,amssymb}
\usepackage{multirow}

\newcommand{\beq}{\begin{equation}}
\newcommand{\eeq}{\end{equation}}

\def\rightcontract{\mathop{\hbox{\vrule width0.5pt height6pt%
  \vrule height0.5pt width6pt}}}

\def\car{_{\rm(car)}}
\def\ucar{u\car}
\def\carrad{_{\rm(rad)}}
\def\carang{_{\rm(ang)}}
\def\rad{{\rm(rad)}}

\def\hr{{\hat r}}
\def\htheta{{\hat \theta}}
\def\hphi{{\hat \phi}}

\begin{document}

\title{Gyroscope precession along bound equatorial plane orbits around a Kerr black hole}

\author{Donato \surname{Bini}$^1$}
\author{Andrea \surname{Geralico}$^2$}
\author{Robert T. \surname{Jantzen}$^{3}$}

\affiliation{
$^1$Istituto per le Applicazioni del Calcolo ``M. Picone'', CNR, I-00185 Rome, Italy\\
$^2$Astrophysical Observatory of Torino, INAF,
I-10025 Pino Torinese (TO), Italy \\
$^3$Department of Mathematics and Statistics, Villanova University, Villanova, PA 19085, USA
}

\date{\today}

\begin{abstract}
The precession of a test gyroscope along stable bound equatorial plane orbits around a Kerr black hole is analyzed and the  precession angular velocity of the gyro's parallel transported spin vector and the increment in precession angle after one orbital period is evaluated. The parallel transported  Marck frame which enters this discussion is shown to have an elegant  geometrical explanation in terms of the electric and magnetic parts of the Killing-Yano 2-form and a Wigner rotation effect. 
\end{abstract}

\maketitle

\section{Introduction}

The precession of the spin of a test gyroscope in a given gravitational field has been studied in great depth in order to open new windows into viable tests of general relativity.
First considered in the pioneering work of Schiff \cite{schiff}, 
this effect has been popularized by the well-known NASA \lq\lq Gravity Probe B" (GP-B) experiment inspired by Schiff. This satellite-based space mission was finally launched in 2004 (with a space-flight phase of about one year) after a preparation of more than forty years.
Two independent contributions are responsible for the gyro precession in geodesic motion, the geodetic effect and the frame-dragging effect.  
The final reported analysis of the data
in the GP-B experiment resulted in a geodetic drift rate of $-6601.8\pm18.3$ mas/yr and a frame-dragging drift rate of $-37.2\pm7.2$ mas/yr, in good agreement with the general relativistic predictions of $-6606.10\pm 0.28$ mas/yr and $-39.20\pm0.19$ mas/yr, respectively  \cite{gpb1,gpb2}. 

Theoretical investigations of test gyroscope spin precession have involved mainly black hole spacetimes (Schwarzschild and Kerr) and  gyroscopes in circular orbits confined to motion in either geodesic or accelerated orbits, primarily in the equatorial  plane \cite{RindlerPerlick:1990,Iyer:1993qa,Bini:1997ea,Bini:1997eb}, the acceleration contributing a Thomas precession term to the total spin precession 
\cite{Jantzen:1992rg,Bini:1994}.
However, there is considerable research on binary systems consisting of two spinning bodies, treated using all the various general relativistic approximation schemes available today, namely post-Newtonian theory \cite{Blanchet:2013haa,Damour:2015isa}, perturbation theory \cite{Sasaki:2003xr}, effective field theories \cite{Levi:2015ixa}, effective-one-body formalism \cite{Buonanno:1998gg,Buonanno:2000ef,Damour:2014sva,Babak:2016tgq}, etc.
Spinning bodies in general relativity can be treated either as pointlike test particles or as extended bodies.
In the former case, the spin direction of a test gyroscope is well known to undergo Fermi-Walker transport along its world line, which for geodesic motion reduces to parallel transport. In the latter case one has instead the so-called Mathisson-Papapetrou-Dixon model \cite{Mathisson:1937zz,Papapetrou:1951pa,Dixon:1970zza} for the evolution of the both the \lq\lq central" world line and spin direction which under certain conditions reduces to the test gyroscope case in the test particle limit when the spin of the object is very small in comparison with its mass. 

Here we consider the case of a test gyroscope in geodesic motion along a periodic bound equatorial orbit in the Kerr spacetime, essentially a \lq\lq precessing ellipse." These orbits allow the generalization of well known results for circular orbits in black hole spacetimes to 
planar orbits of
nonzero eccentricity.
Indeed, in a stationary axisymmetric asymptotically flat spacetime one has a local coordinate grid  
which is rigidly connected to radial infinity and provides a way to measure the local precession of the spin direction with respect to some fixed Cartesian frame at infinity. By considering planar motion in the equatorial plane of a black hole spacetime, the situation is much simpler to discuss, since the precession is confined to a single angle in 2 spatial dimensions.

The natural spherical orthonormal frame associated with the static Killing observers moving along the Boyer-Lindquist coordinate time lines in the Kerr spacetime has axes which are tied to the directions of incoming photons from the distant stars at fixed angular locations at spatial infinity in the Boyer-Lindquist coordinate grid. A test gyroscope moves relative to these observers causing a kinematical deformation of its static observer measured spin vector due to stellar aberration. This can be eliminated by boosting the static observer axes to the local rest space of the gyro along its orbit, leading to a formula for the spin precession angular velocity relative to these boosted axes which has a nice geometrical interpretation in terms of the gravitoelectromagnetic (threading \lq\lq 1$+$3'') decomposition of the gravitational field \cite{Jantzen:1992rg}.

The key to the evaluation of this precession is the Carter observer orthonormal frame \cite{Carter:1968ks} which is intimately associated with the Killing vector and tensor constants of the geodesic motion which in turn lead to the effective potential description of geodesic motion. Starting from the Carter frame, Marck \cite{marck1,marck2} discovered a parallel transported frame adapted to the local rest space of timelike geodesics in the Kerr spacetime which allows the precession  to be evaluated simply in terms of the constants of the motion, modulo a generalized Wigner rotation effect \cite{Wigner:1939cj} associated with the motion relative to the static observers that, like stellar aberration, does not contribute to the average precession. This ``spin aberration" effect due to the generalized Wigner rotation is the result of three successive boosts required to pass through the Carter frame on the way to the local rest space of the gyroscope starting from the static observer frame, and naturally emerges from the geometry underlying the Marck frame. Although this discussion applies to general geodesic motion, we only investigate it here for equatorial plane orbits, providing explicit expressions for frames, precession frequencies, and the accumulated spin rotation angle after an azimuthal period of the motion.

\section{Bound equatorial plane orbits around a Kerr black hole}

Consider the Kerr metric written in standard Boyer-Lindquist coordinates $(x^\alpha)=(t,r,\theta,\phi)$
\begin{eqnarray}\label{K1}
d s^2 &=& g_{\alpha\beta}dx^\alpha dx^\beta\nonumber\\
&=&-d t^2+\frac{\Sigma}{\Delta} \,d r^2+\Sigma \,d\theta^2 +(r^2+a^2)\sin^2\theta \,d\phi^2\nonumber\\
&&
+\frac{2Mr}{\Sigma}(dt-a\sin^2\theta \,d\phi)^2\,,
\end{eqnarray}
where $a=J/M$ is the specific angular momentum of the source (with $\hat a=a/M$ dimensionless)  and
\beq\label{K2}
\Sigma=r^2+a^2\cos^2\theta\,,\qquad \Delta=r^2-2Mr+a^2\,.
\eeq
The outer horizon radius is at $r_+=M+\sqrt{M^2-a^2}$.
Units are chosen here such that $G=c=1$. The static observers move along the time coordinate lines with 4-velocity $m=(-g_{tt})^{-1/2}\, \partial_t$ aligned with the Killing vector field $\partial_t$ and play a fundamental role in the spin precession as seen by observers far from the black hole. 

Timelike geodesic world lines in this metric  $x^\alpha=x^\alpha (\tau)$ parametrized by the proper time $\tau$ have a 4-velocity $U^\alpha=dx^\alpha/d\tau$ whose coordinate components satisfy
\begin{eqnarray}
\label{geosgen}
\frac{d t}{d \tau}&=& \frac{1}{\Sigma}\left[aB+\frac{(r^2+a^2)}{\Delta}P\right]\,,\nonumber \\
\frac{d r}{d \tau}&=&\epsilon_r \frac{1}{\Sigma}\sqrt{R}\,,\nonumber \\
\frac{d \theta}{d \tau}&=&\epsilon_\theta \frac{1}{\Sigma}\sqrt{\Theta}\,,\nonumber \\
\frac{d \phi}{d \tau}&=& \frac{1}{\Sigma}\left[\frac{B}{\sin^2\theta}+\frac{a}{\Delta}P\right]\,,
\end{eqnarray}
where $\epsilon_r$ and $\epsilon_\theta$ are sign indicators, and
\begin{eqnarray}
\label{geodefs}
P&=& E(r^2+a^2)-La\,,\nonumber\\
B&=& L-aE \sin^2\theta\,, \nonumber\\
R&=& P^2-\Delta (r^2+K)\,,\nonumber\\
\Theta&=&K-a^2\cos^2\theta-\frac{B^2}{\sin^2\theta}\,,
\end{eqnarray}
where $K$ is Carter's constant  associated with the symmetric Killing 2-tensor 
of the Kerr spacetime~\cite{marck1,marck2}
and $E$ and $L$ are the conserved energy and angular momentum per unit mass
associated with the Killing vector fields $\partial_\phi$ and $\partial_t$ of a test particle in geodesic motion.
Note that $E$ and $L/M$ are dimensionless. 

We are interested here in equatorial orbits, i.e., orbits at $\theta=\pi/2$ with $K=(L-a\,E)^2=x^2$ (with $\hat x=x/M$ dimensionless) so that
\begin{eqnarray}
\label{eq_t}
\Delta\,r^2 \,\frac{dt}{d\tau} &=& 
(Er^2-ax)(r^2+a^2)+ \Delta a x \\
\label{eq_r}
r^4 \,\Big(\frac{dr}{d\tau}\Big)^2  &=& [r^2 E -a x]^2-\Delta(r^2+x^2)\,, \\
\label{eq_phi}
\Delta\,r  \,\frac{d\phi}{d\tau}&=& 
rL-2Mx\,.
\end{eqnarray} 
The vertical direction along $\partial_\theta$ at the equatorial plane is covariant constant there, and the precession of a test gyroscope in such an orbit only undergoes a rotation in the 2-plane of the radial and azimuthal directions. These directions are locked to the observers at rest at spatial infinity, and so provide a natural way to measure the spin precession as seen by distant observers, modulo the boost between the local rest space of the gyro and that of the static observers tied to the coordinate grid.

We limit our considerations to bound orbits ($0<E<1$) which oscillate between a minimum radius $r_{\rm per}$ (periastron) and a maximum radius $r_{\rm apo}$ (apastron), namely periodic motion at the period of the radial motion. The points on such an orbit corresponding to these extremal radii precess since the period of the azimuthal motion is distinct from that of the radial motion.
For nonzero eccentricity, the radial variable along these  precessing ellipses can be expressed in the form
\beq\label{rversuschi}
r =\frac{Mp}{1+e\cos \chi}\,,
\eeq
where $\chi$ is a new function of the proper time along world line of the gyro. The extremal values of the radii are then 
\beq\label{rextrema}
   r_{\rm per} = \frac{M p}{1+e}\,,\qquad
  r_{\rm apo} = \frac{M p}{1-e} \,,
\eeq
in terms of which one can express the eccentricity $0\le e<1$ and semi-latus rectum $Mp$ of these precessing ellipses
\begin{eqnarray}\label{epdef}
e&=& \frac{r_{\rm apo}-r_{\rm per}}{r_{\rm apo}+r_{\rm per}}\,,\qquad
Mp = \frac{2r_{\rm per}\,r_{\rm apo}}{r_{\rm per}+r_{\rm apo}}\,.
\end{eqnarray}
Note that $p$ is dimensionless, as is its reciprocal $u_p=1/p$.

We assume $a\ge 0$ in order to define prograde (corotating) and retrograde (counterrotating) orbits by the signs $+$ and $-$ respectively of the nonzero azimuthal angular velocity  $d\phi/d\tau$ or equivalently of the angular momentum $L$. Formulas valid for retrograde orbits can be obtained from those for prograde orbits by $a\to -a$ and $L \to - L$, under which $x\to -x$.

Eq.~\eqref{eq_r} can be rewritten in  factorized form
\beq
\label{r_eq2}
\left(\frac{dr}{d\tau}\right)^2=-\frac{(1-E^2)}{r^3}(r-r_3)(r-r_{\rm apo})(r-r_{\rm per})\,,
\eeq
where
\beq
\label{eq_ni_r3EX}
\frac{r_3}{M} = \frac{2\hat x{}^2(1-e^2)}{p^2(1- E^2)}\,. 
\eeq
The motion is confined to $r_{\rm per}\le r \le r_{\rm apo}$, which therefore requires the third root to satisfy
$r_3 < r_{\rm per}$. In fact when $r_3 = r_{\rm per}$, the effective potential for radial motion has a critical point with a negative second derivative at the periastron corresponding to an unstable circular orbit radius $r_c$, making the eccentric orbit at that energy marginally stable \cite{Glampedakis:2002ya}.
This condition on allowed values of $(e,p)$ determines the \lq\lq separatrix" of the bound orbits, whose parametric equations are given by \cite{Levin:2008yp}
\begin{eqnarray}
e^{\rm sep}&=&-\frac{r_c^2-6Mr_c-3a^2\pm8a\sqrt{Mr_c}}{\Delta_c}\,, \nonumber\\
p^{\rm sep}&=&\frac{4r_c}{\Delta_c}(\sqrt{Mr_c}\mp a)^2\,,
\end{eqnarray}
with $\Delta_c=\Delta(r_c)$.
These may be re-expressed in terms of the parameter $u_p=1/p$ using $u_p(1+e)=M/r_c$ following from \eqref{epdef} with $r_{\rm per}=r_c$ to get the terminal values needed below of functions of $u_p$ at the marginally stable bound orbits.

Using \eqref{rextrema} which expresses $(r_{\rm per},r_{\rm apo})$ in terms of $(e,p)$,
the two conditions
\beq\label{rextrema}
\left(\frac{dr}{d\tau}\right)\Big|_{r_{\rm per}}=0=\left(\frac{dr}{d\tau}\right)\Big|_{r_{\rm apo}}\,,
\eeq
can be imposed on Eq.~\eqref{eq_r} to solve them for $E=E(p,e)$ and $L=L(p,e)$ as follows. 
Then expand the extremal conditions \eqref{rextrema} as functions of $p,e$ and subtract them to identify $E^2$ as
\beq\label{Epex}
E^2 =\frac{1}{p}\left[ \frac{(1-e^2)^2  \hat x{}^2}{p^2}+p-(1-e^2) \right] \,.
\eeq
Backsubstituting this into either of these conditions leads to the quadratic equation
\beq
\hat x{}^2 +\frac{2  \hat a E p}{(p-3-e^2)} \hat x -\frac{p(p  -\hat a^2)}{(p-3-e^2)}=0 \,.
\eeq

Solving this final equation  
for $ E$ as a function of $\hat x$
\begin{eqnarray}
 E= -\frac{p-3-e^2}{2\hat a   p} \hat x 
-\frac{ (\hat a^2-p)}{2\hat a }\frac{1}{\hat x} \,,
\end{eqnarray}
and then substituting this expression for $E$ into  Eq.~\eqref{Epex} one obtains
a  quartic equation for $\hat x$
\beq
F \hat x{}^4 +N \hat x{}^2 +C=0\,,
\eeq
with dimensionless coefficients $F$, $N$ and $C$ given by \cite{Glampedakis:2002ya}
\begin{eqnarray}
 F  &=&  \left(1-\frac{ 3+e^2 }{p} \right)^2-\frac{4\hat a^2(1-e^2)^2}{p^3}  \,,\nonumber\\
-\frac{N}{2}&=&(p-3-e^2)+\hat a^2\left(1 +\frac{1+3e^2}{p}\right)\,, \nonumber\\
C&=&(\hat a^2-p)^2\,.
\end{eqnarray}
The solution  is then  
\beq\label{x2sol}
\hat x{}^2 =\frac{-N\mp \sqrt{N^2-4CF}}{2F}\,, 
\eeq
where the upper (lower) sign corresponds to prograde (retrograde) motion and
\begin{eqnarray}\label{discriminant}
N^2-4CF&=&\frac{16\hat a^2}{p^3}\{[p^2-2p+\hat a^2(1+e^2)]^2\nonumber\\
&&
-4e^2(p-\hat a^2)^2]\}\,.
\end{eqnarray}

To understand this last sign correlation, consider the absolute value $|x|=|L-aE|$. If $L$ and $a$ are both the same (opposite) sign, we have a prograde (retrograde) orbit, and so clearly $|x_{\rm pro}| < |x_{\rm retro}|$. Assuming $a>0$, this requires that $x_{\rm pro}>0$ and $x_{\rm retro}<0$. Stable circular orbits have $N<0$, so the minus sign in Eq.~\eqref{x2sol} gives the smaller root in absolute value and must correspond to the prograde orbit, so the positive square root is relevant and must be chosen. Similarly the positive sign gives the larger root in absolute value so must correspond to the retrograde orbit, so the negative square root value is relevant and must be chosen, resulting in equations with $\mp$ to distinguish the prograde and retrograde orbits respectively. This sign is directly correlated with the sign $\mp =-{\rm sgn}(a)$, so formally one can make this replacement and combine it with the overall factor of $|\hat a|$ one can factor out of the square root of the discriminant \eqref{discriminant} to have a factor of $\hat a$ in front of that square root in the solution, and changing the sign of $a$ will then correctly interchange these two physical roots.

The explicit expressions for $E,L,\hat x$ expanded in a series in $e^2$ up to first order for prograde orbits are

\begin{widetext}

\begin{eqnarray}
E&=&\frac{1-2u_p+{\hat a}u_p^{3/2}}{\sqrt{1-3u_p+2{\hat a}u_p^{3/2}}}\left[
1+
\frac{-2{\hat a}^4u_p^4+3{\hat a}^3u_p^{7/2}+u_p^2(-1+10u_p){\hat a}^2-u_p^{3/2}(-7+26u_p){\hat a}+(4u_p-1)^2}{2(1-2u_p+{\hat a}u_p^{3/2})(1-3u_p+2{\hat a}u_p^{3/2})(1-2u_p+{\hat a}^2u_p^2)}u_pe^2\right]
\,,\nonumber\\
\frac{L}{M}&=&\frac{1-2{\hat a}u_p^{3/2}+{\hat a}^2u_p^2}{\sqrt{u_p(1-3u_p+2{\hat a}u_p^{3/2})}}\left\{
1-
\left[
\frac12
+\frac{{\hat a}u_p^{1/2}(1+u_p)}{1-2u_p+{\hat a}^2u_p^2}
+\frac{1-4u_p}{2(1-3u_p+2{\hat a}u_p^{3/2})}
-\frac{1+{\hat a}u_p^{1/2}(1-u_p)}{1-2{\hat a}u_p^{3/2}+{\hat a}^2u_p^2}
\right]e^2
\right\}
\,,\nonumber\\
\hat x&=&\frac{1-{\hat a}u_p^{1/2}}{\sqrt{u_p(1-3u_p+2{\hat a}u_p^{3/2})}}\left\{
1-
\left[
-\frac12
+\frac{2{\hat a}u_p^{3/2}}{1-2u_p+{\hat a}^2u_p^2}
+\frac{1-4u_p}{2(1-3u_p+2{\hat a}u_p^{3/2})}
\right]e^2
\right\}
\,.\nonumber\\
\end{eqnarray}

Consider now the radial equation \eqref{eq_r} and use the relation \eqref{rversuschi} to introduce the angular variable $\chi$ in place of $r$ along a given orbit (for $e>0$). One finds
\beq\label{dphidchi}
\frac{d\phi}{d\chi}= u_p^{1/2}\frac{ \hat x + \hat a E - 2 u_p \hat x (1+  e\cos \chi) }{[1+u_p^2 \,\hat x{}^2 (e^2-2 e\cos \chi -3) ]^{1/2}
[1-2 u_p(1+ e\cos \chi) +a^2 u_p^2(1+ e\cos \chi)^2  ]}
\eeq
and
\begin{eqnarray}
M \frac{d\chi}{d\tau} &=&  u_p^{3/2}(1+e\cos \chi )^2
[1+\hat x{}^2 u_p^2 ( e^2-2 e\cos\chi-3)]^{1/2}\nonumber
\end{eqnarray}
together with 
\begin{eqnarray}
\frac{dt}{d\tau} &=&
\frac{D_1}{D_2}\,\nonumber\\
M \frac{d\chi}{dt} &=&  u_p^{3/2}(u_p^2\hat x^2 e^2 -3 u_p^2\hat x^2+1-2 u_p^2\hat x^2 e\cos \chi)^{1/2} (1+e\cos \chi )^2
\,\frac{D_2}{D_1}
\,,
\end{eqnarray}
where
\begin{eqnarray}
D_1&=& E +E\hat a^2 u_p^2 (1+e\cos\chi)^2   -2 \hat a u_p^3\hat x (1+e\cos\chi)^3 \,,\nonumber\\
D_2&=& 1-2 u_p(1+ e\cos \chi) +a^2 u_p^2(1+ e\cos \chi)^2  \,.
\end{eqnarray}

\end{widetext}
Similar relations for expressing  derivatives of related quantities  as functions of $\chi$ are easily found.

Note that in the limit of zero eccentricity, Eq.~\eqref{dphidchi} becomes for prograde orbits
\beq
  \left. \frac{d\phi}{d\chi} \right|_{e=0}
= (1-6u_p +8a u_p^{3/2}-3a^2 u_p^2)^{- 1/2} \,,
\eeq
which is the Kerr azimuthal to radial epicyclic frequency ratio determining the precession of the almost circular orbits during one coordinate time period of the radial motion, i.e., the rotation of the periastron of the orbit. Indeed the integral of Eq.~\eqref{dphidchi} gives the increment of precession of the periastron in azimuthal angle for any eccentricity during one radial period (modulo $2\pi$)
\beq
  \Delta\phi_{\rm orb} = \int_0^{2\pi} \frac{d\phi}{d\chi} \, d\chi  -2\pi\,{\rm sgn}\left(\frac{d\phi}{d\chi}\right)\,,
\eeq
where the radial period and corresponding radial frequency are
\beq
T_r = \int_0^{2\pi} \frac{dt}{d\chi} \, d\chi\,,\qquad
\Omega_r = \frac{2\pi}{T_r}
\,.
\eeq
The constant rate of precession of the periastron is then the ratio  $\Delta\phi_{\rm orb}/T_r$.
Similarly the azimuthal frequency is
\beq
\Omega_\phi = \frac1{T_r}\int_0^{2\pi} \frac{d\phi}{d\chi} \, d\chi\,.
\eeq
Both frequencies can be expressed in terms of elliptic integrals.
Explicitly the (dimensionless) coordinate time orbital frequencies of the radial and azimuthal motions are respectively up to order $O(e^2)$ given by

\begin{widetext}

\begin{eqnarray} \label{frequencies}
M \Omega_{r}&=&
\frac{u_p^{3/2}\sqrt{1-6u_p+8{\hat a}u_p^{3/2}-3{\hat a}^2u_p^2}}{1+{\hat a}u_p^{3/2}}\left\{
1-\frac34\frac1{(1+{\hat a}u_p^{3/2})(1-2u_p+{\hat a}^2u_p^2)(1-6u_p+8{\hat a}u_p^{3/2}-3{\hat a}^2u_p^2)^2}\right.\nonumber\\
&&
\times 
[2-266u_p^3-32u_p+165u_p^2-u_p^{3/2}(376u_p-841u_p^2+2u_p^3-38){\hat a}-u_p^2(12-314u_p+999u_p^2+16u_p^3){\hat a}^2\nonumber\\
&&
+u_p^{7/2}(-108+466u_p+93u_p^2){\hat a}^3-u_p^4(-11-32u_p+176u_p^2){\hat a}^4+u_p^{11/2}(-101+160u_p){\hat a}^5\nonumber\\
&&\left.
-u_p^6(-25+72u_p){\hat a}^6+13u_p^{15/2}{\hat a}^7]
e^2\right\}
+O(e^3)\,,\nonumber\\
M \Omega_{\phi}&=&
\frac{u_p^{3/2}}{1+{\hat a}u_p^{3/2}}\left[
1-3\frac{1+2u_p(-5+11u_p)-u_p^{3/2}(-11+42u_p){\hat a}+3u_p^2(-1+8u_p){\hat a}^2-u_p^{7/2}{\hat a}^3-2u_p^4{\hat a}^4}{2(1+{\hat a}u_p^{3/2})(1-2u_p+{\hat a}^2u_p^2)(1-6u_p+8{\hat a}u_p^{3/2}-3{\hat a}^2u_p^2)}e^2\right]\nonumber\\
&&
+O(e^3)\,.
\end{eqnarray}

Fig.~\ref{fig:1} plots these expressions using the exact formulas and not approximate ones expanded in terms of the eccentricity. The same is true of Fig.~\ref{fig:2}.


\begin{figure*}
\centering
\includegraphics[scale=0.4]{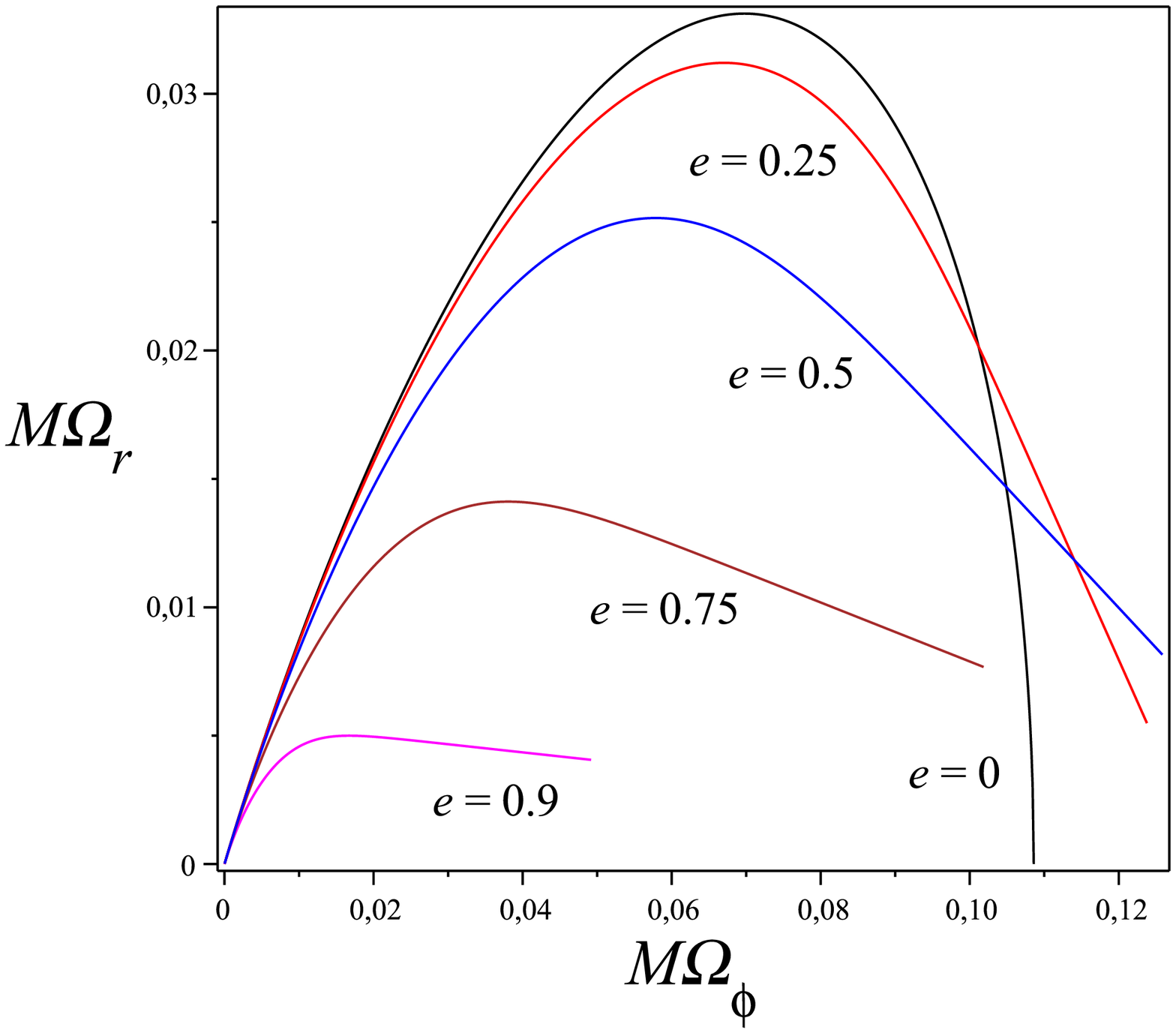}
\includegraphics[scale=0.4]{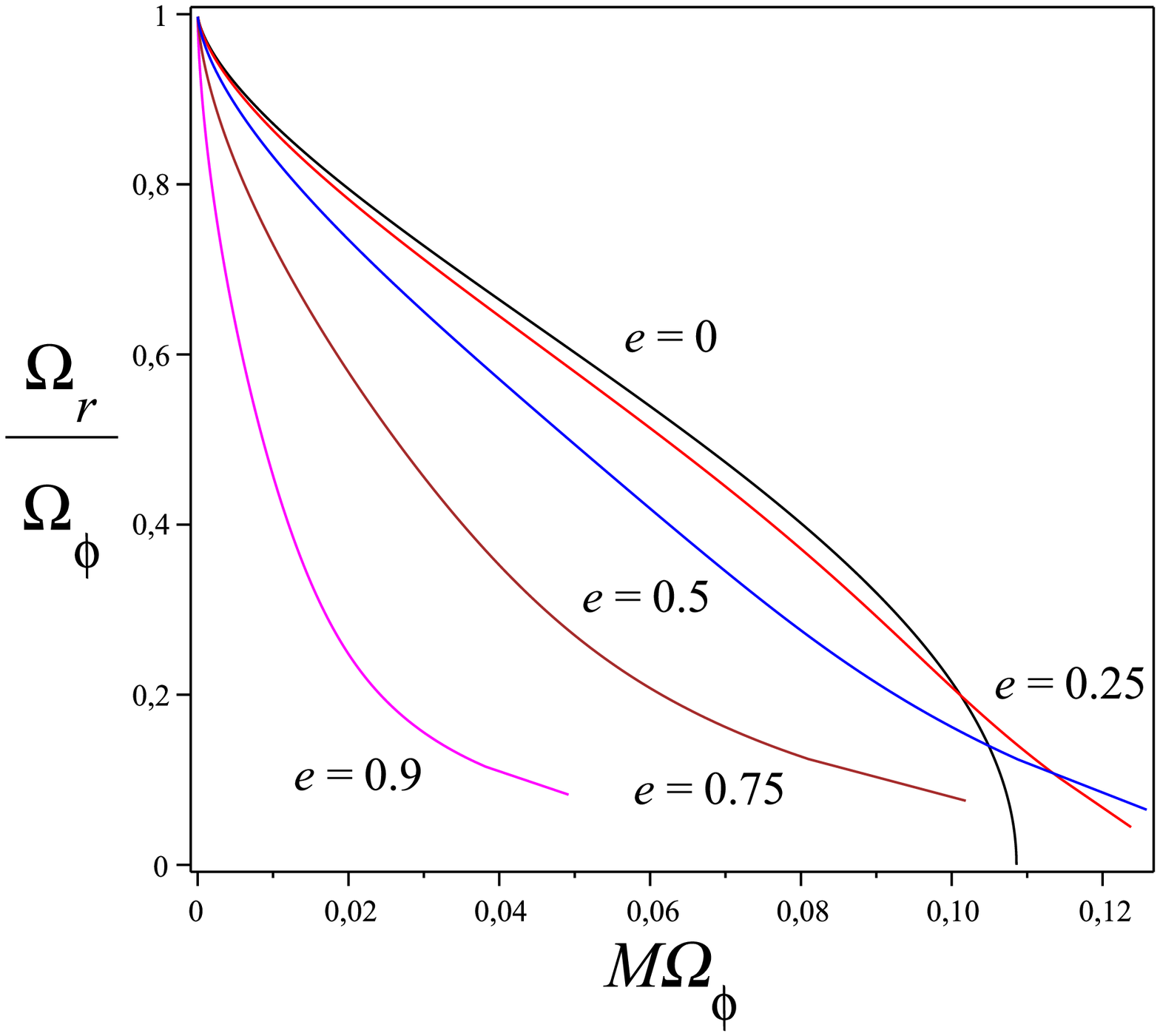}
\caption{The behavior of the dimensionless orbital frequencies $M\Omega_{r}$ and $M\Omega_{\phi}$ versus each other (left panel) and their ratio  versus $M\Omega_{\phi}$ (right panel) is shown for $a=0.5 M$ and selected values of the eccentricity $e$. The plots are made parametrically in $u_p$ ranging from $u_p=0$ (at the left endpoints corresponding to radial infinity where both frequencies approach each other and zero) to the eccentricity-dependent values of $u$ corresponding to the marginally stable orbits  for corotating geodesics. Only the circular orbit case ($e=0$) reaches the horizontal axis.
\label{fig:1}}
\end{figure*}

Finally the equations of motion can be fully integrated in terms of special functions, but the corresponding expressions are not very illuminating. For simplicity following \cite{Glampedakis:2002ya}, we discuss the eccentricity corrections to circular motion up to  order $e^2$. 
Using the coordinate time as a parameter we find explicitly
\begin{eqnarray}
\label{rdit}
\frac{r(t)}{ M}&=&
R_0 +e \, R_1(\cos (\Omega_{r} t)-1)+e^2 \, R_2(\cos(2\Omega_{r} t)-1)
+O(e^3)\,,\\
\label{phidit}
\phi(t) &=&  \Omega_{\phi} t+ e \, \Phi_1\sin(\Omega_{r} t)+e^2 \, \Phi_2\sin(2\Omega_{r} t) 
+O(e^3)\,,
\end{eqnarray}
where
\begin{eqnarray}
R_0 &=&  \frac{1+e+e^2}{u_p}\,,\qquad 
R_1 = \frac{1}{u_p}\,,\nonumber\\
\frac{R_2}{R_1}&=&-\frac{2}{3} +\frac{ u_p(1+3u_p)}{(1+{\hat a}u_p^{3/2})(1+u_p-2u_p^2)}
+\frac{2u_p^{5/2}[2u_p^{1/2}-{\hat a}(1+u_p)]}{(1-2 u_p+{\hat a}^2u_p^2)(1+2 u_p)(1- u_p)}
+\frac1{6}\frac{1-3 u_p+2{\hat a}u_p^{3/2}}{1-6u_p+8{\hat a}u_p^{3/2}-3{\hat a}^2u_p^2}
\,,\nonumber\\
\Phi_1 &=& -2\frac{1-3 u_p+2{\hat a}u_p^{3/2}}{\sqrt{1-6u_p+8{\hat a}u_p^{3/2}-3{\hat a}^2u_p^2}(1+{\hat a}u_p^{3/2})(1-2 u_p+{\hat a}^2u_p^2)}
\,,\nonumber\\
\frac{\Phi_2}{\Phi_1} &=& 
-\frac{19}{48} +\frac{ u_p(1+3u_p)}{(1+{\hat a}u_p^{3/2})(1+u_p-2u_p^2)}
-\frac12\frac{1-3u_p^2(1+2 u_p)+4{\hat a}u_p^{3/2}(1+u_p)}{(1-2 u_p+{\hat a}^2u_p^2)(1+2 u_p)(1- u_p)}
+\frac{3}{16}\frac{1+ u_p}{1-3 u_p+2{\hat a}u_p^{3/2}}\nonumber\\
&&
+\frac1{12}\frac{1-3 u_p+2{\hat a}u_p^{3/2}}{1-6u_p+8{\hat a}u_p^{3/2}-3{\hat a}^2u_p^2}
\,.
\end{eqnarray}

\end{widetext}

\section{Marck's parallel propagated frame as a Frenet-Serret frame}

Marck completed the gyro 4-velocity $U=e_0$ along an arbitrary geodesic in the Kerr spacetime to a parallel transported frame \cite{marck2} using Kerr's Killing-Yano tensor 2-form $f$  whose nonvanishing (coordinate) components are given by
\begin{eqnarray}
\label{KY}
f_{tr}&=&-a\cos\theta\,,\qquad f_{t\theta}=ar \sin \theta \,,\nonumber\\ 
f_{r\phi}&=&-a^2\cos \theta \sin^2 \theta \,,\qquad f_{\theta\phi}=(r^2+a^2)r\sin \theta \,,\nonumber\\
\end{eqnarray}
and which satisfies
$\nabla_{(\rho} f_{\nu)\mu} =0$. The Killing tensor $K_{\alpha\beta}=f_\alpha{}^\gamma f_{\gamma\beta}$ leads to Carter's constant of the motion $K=K_{\alpha\beta}U^\alpha U^\beta$. 
In the equatorial plane the second frame vector is then obtained by forming the unit spacelike 1-form 
\begin{equation}\label{K29}
e_\mu{}_{2}=\frac{1}{x}f_{\mu\nu} U^\nu =  r \delta^\theta{}_\mu
\end{equation}
which is orthogonal to $e_{0}$  and is parallel propagated along the geodesic orbit.  
Marck then completed this to an orthonormal frame by adding the two vector fields whose corresponding 1-forms (indicated by $\flat$) at the equatorial plane are \cite{Bini:2016xqg} 
\begin{eqnarray}
   \tilde e_{1} {}^\flat
&=& \frac{x}{\sqrt{x^2+r^2}} \left[ -r \dot r \, (dt-a\,d\phi)+\frac{r}{\Delta} (r^2 E-a x)\,dr\right] \,,\nonumber\\
   \tilde e_{3} {}^\flat
 &=& \frac{x}{\sqrt{x^2+r^2}} \left[
\, \frac{r^2 }{\Delta}\,\dot r dr  -
\, \frac{(r^2 E-a x)}{r^2} (dt -a d\phi) \right]\nonumber\\
&&
- \frac{\sqrt{x^2+r^2}}{r^2} 
\,  [a\, dt -(r^2+a^2)\, d\phi] \,,
\end{eqnarray}
where $\dot r = dr/d\tau$ is given by \eqref{eq_r}.

This frame is a degenerate  Frenet-Serret frame along the geodesic
\begin{eqnarray}
\frac{De_{0}}{d\tau} &=& 0\,,\qquad \frac{De_{2}}{d\tau} = 0\,,\nonumber\\
\frac{D\tilde e_{1}}{d\tau} &=&{\mathcal T} \tilde e_{3}  \,,\qquad 
\frac{D\tilde e_{3}}{d\tau} = -{\mathcal T} \tilde e_{1}\,,
\end{eqnarray}
with Frenet-Serret rotation vector $\omega_{\rm(FS)}=-{\mathcal T} e_{2}$.
Appendix A discusses how these last two frame vector fields come about, corresponding respectively to the radial and azimuthal directions in the local rest space of the geodesic.
Rotating them by a clockwise rotation angle $\Psi$ in the $\tilde e_{1}$-$\tilde e_{3}$ plane
to get a parallel propagated frame
\beq\label{K30}
\left(\begin{array}{c} e_1 \\ e_3 \end{array}\right)
= R(\Psi) \left(\begin{array}{c} \tilde e_1 \\ \tilde e_3 \end{array}\right)
\equiv\left(\begin{array}{cc}  \cos \Psi &- \sin \Psi \\ \sin \Psi & \cos \Psi\end{array}\right)
\left(\begin{array}{c} \tilde e_1 \\ \tilde e_3 \end{array}\right)
\,,
\eeq
one finds the angular velocity ${\mathcal T}$ of the gyro-fixed axes with respect to the preliminary Marck axes (in the clockwise direction) \cite{marck1,marck2}
\beq\label{K31bis}
 {\mathcal T} = \frac{d\Psi}{d\tau}
=\frac{a+E x}{r^2+x^2}\,.
\eeq
For a circular orbit at constant $r$, this is then a constant leading to a uniform rotation of the spin vector.

A direct evaluation of ${\mathcal T}$ expanded to second order in the eccentricity $e$ yields

\begin{widetext}

\begin{eqnarray}
 M {\mathcal T} &=& u_p^{3/2}\left\{
1+
2\frac{1-3u_p+2{\hat a}u_p^{3/2}}{1-2u_p+{\hat a}^2u_p^2}e\cos\chi\right.\nonumber\\
&&\left.
+\left[
\frac{(1-3u_p+2{\hat a}u_p^{3/2})(1-6u_p+8{\hat a}u_p^{3/2}-3{\hat a}^2u_p^2)\cos^2\chi}{(1-2u_p+{\hat a}^2u_p^2)^2}
+\frac{u_p(1-{\hat a}u_p^{1/2})^2(1-4u_p+4{\hat a}u_p^{3/2}-{\hat a}^2u_p^2)}{(1-2u_p+{\hat a}^2u_p^2)^2}
\right]e^2
\right\}\nonumber\\
&&
+O(e^3)
\,.
\end{eqnarray}

\end{widetext}
This corrects the Kerr circular orbit value for small eccentricity, recalling $u_p=M/r_c$ for circular orbits.

\section{The boosted spherical frame}

The fact that the motion is planar makes it easier to understand the precession of the two planar gyro-fixed axes (radial and azimuthal directions) along a geodesic compared to such axes in a fixed Cartesian frame at spatial infinity, since the tilting in time to map to the local rest space is irrelevant to the rotation of these two directions, apart from proper time considerations and some deviational rotational behavior which averages to zero over a radial period of the motion. The spherical coordinate grid is seen by observers at radial infinity as nonrotating so by measuring the spin relative to this grid, we can evaluate how an observer at radial infinity sees the spin direction change. The directions of incoming photons from the \lq\lq fixed stars" are locked to the axes associated with the static observers, and boosting these axes to the local rest space of the orbiting gyroscope removes the effect of stellar aberration which does not contribute to the average precession.
If at an initial point on an orbit we fix an orthonormal triad of vectors aligned with the static observer axes boosted into the local rest space of the gyroscope along the azimuthal direction, we can simply rotate them by the opposite signed increment of the azimuthal coordinate $\phi$ with respect to the static observer axes along the orbit to keep their direction \lq\lq fixed" with respect to radial infinity. This defines a \lq\lq nonrotating" static frame whose axes then realign with the static observer spherical axes each time the orbit returns to the same value of the coordinate $\phi$ \cite{Jantzen:1992rg}.

The spin vector along the orbit projected onto the sequence of static observer axes does not rotate simply with respect to any orthonormal triad adapted to the static observer local rest spaces along the orbit but undergoes a periodic distortion away from a simple rotation both in magnitude and direction due its projection onto the static observer axes along its orbit
\cite{Jantzen:1992rg}, similar to the spin vector of a classical electron in a circular orbit undergoing Thomas precession \cite{Misner:1974qy,Bini:2002mh}. On the other hand, by boosting the spin vector from the local rest frame of the geodesic to the static observer local rest space along the geodesic these periodic distortions due to the relative motion are removed. This allows the definition of a simple rotation with respect to the static observer spherical frame with a definite angular velocity of precession, from which one must subtract the angular velocity due to orbital rotation of the spherical frame fixed with respect to spatial infinity.  Alternatively one can boost this \lq\lq nonrotating" static frame into the local rest space of the orbit to compare with the spin vector and evaluate an angular velocity of precession as seen from spatial infinity.
 
The complication comes from the fact that the Marck frame also undergoes a periodic rotation with respect to such a boosted nonrotating static frame, but one which averages out to the identity, in the same way that the distortions in the measured spin vector by the sequence of static observers is not relevant to the spin precession. This rotation can be calculated (see Appendix C) but over a radial period of the motion it does not contribute to the net precession of the spin vector and will be ignored here.
 
The Marck frame vector $\tilde e_{1}$ is locked to the radial direction $e_{\hat r} $  in the spherical grid of the static observers following the time lines, 
differing only but a boost due to the radial motion of the gyro alone, not to the boost of the relative motion (see Appendix C).
This grid does not rotate with respect to observers at rest at spatial infinity. Along the geodesics $e_{\hat r} $ rotates with respect to fixed Cartesian axes at radial infinity by a rate determined by the orbital angular velocity $d\phi/d\tau$ measuring the rate of rotation of these axes in the counterclockwise direction of increasing $\phi$ coordinate. Subtracting the angular velocity  ${\mathcal T}$ of the gyro axes in the clockwise direction gives the total coordinate time angular velocity of the gyro spin relative to axes whose directions are fixed with respect to radial infinity as
\beq\label{omegaprec}
  \Omega_{\rm prec} = \frac{d\tau}{dt} \left( \frac{d\phi}{d\tau} -\mathcal T   \right)
  = \frac{d}{dt} \left(\phi-\Psi  \right)
\,.
\eeq
This corresponds to the clockwise rotation by $\phi$ of some initial spherical axes in the rest frame of the orbit, choosing $\Psi=0$ at $\phi=0$ to align them with the spherical frame vectors initially at $\tau=0$,  which are then rotated counterclockwise by the angle $\Psi$ to keep them \lq\lq parallel" to the original axes (in the sense of parallel transport)
\beq
   \left(\begin{array}{c} e_{1}\\e_{3} \end{array} \right)
= \left(\begin{array}{cc} \cos(\phi-\Psi)&  \sin(\phi-\Psi)\\ -\sin(\phi-\Psi)&  \cos(\phi-\Psi) \end{array} \right)
   \left(\begin{array}{c}  e_{1}(0)\\  e_{3}(0) \end{array} \right)
\,.
\eeq
It is the difference between these two opposing angles which leads to a net precession after one azimuthal revolution $\phi:0\to\pm2\pi$, by an
increment 
\beq
\Delta\Phi=\Delta(\phi-\Psi)=\pm2\pi-\Psi\vert_{\phi=\pm 2\pi}
\eeq
which represents the advance of the precession angle with respect to direction of the azimuthal motion, with the $\pm$ sign correlated with increasing and decreasing values of $\phi$ along the orbit.

Note that the proper time precession angular velocity has the following simple representation in terms of the constants of the motion
\beq
 \Omega_{\rm prec} \frac{dt}{d\tau} = \frac{L-2Mx/r}{\Delta}-\frac{a+E x}{r^2+x^2}   
\rightarrow \frac{(1-E)L}{r^2}  \,,
\eeq
which has the same sign as $L$ for $a\to0$ and $r/M\to0$ corresponding to the advance of the spin with respect to the orbital motion in the large radius limit of a slowly rotating black hole. 
Indeed  this is always the case for all allowed radii and all values of the orbital parameters for stable bound orbits.

The spin precession angular velocity $\Omega_{\rm prec}$ will be a periodic function of $t$ having the period of the radial motion, and a periodic function of $\chi$ with period $2\pi$. 
When taking the average  value of this quantity over a coordinate time radial period gives

\begin{figure}
\centering
\includegraphics[scale=0.4]{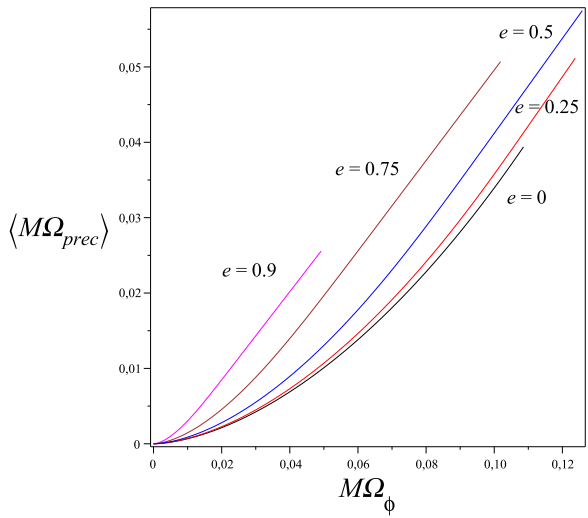}
\caption{The behavior of the average dimensionless precession frequency $\langle M\Omega_{\rm prec}  \rangle$ versus the dimensionless azimuthal orbital frequency $M\Omega_{\phi}$ is shown for $a=0.5 M$ and selected values of the eccentricity $e$ for corotating orbits, as in Fig.~\ref{fig:1}.
\label{fig:2}}
\end{figure}

\begin{widetext}

\begin{eqnarray}
\langle \Omega_{\rm prec}  \rangle &=& \frac{1}{T_r}\int_0^{T_r} \Omega_{\rm prec} dt 
=-\frac{1}{T_r}\int_0^{2\pi} \frac{d\tau}{d\chi}  {\mathcal T} d\chi  
+ \frac{1}{T_r}\int_0^{2\pi} \frac{d\phi}{d\chi} d\chi
\nonumber\\
&=& -
\frac{(\hat a+E \hat x) u_p^{1/2}}{T_r}\int_0^{2\pi} \frac{d\chi }{
[1+\hat x{}^2 u_p^2 ( e^2-2 e\cos\chi-3)]^{1/2}(1+\hat x^2 u_p^2  (1+e\cos \chi)^2)}     + \frac{1}{T_r}\int_0^{2\pi} \frac{d\phi}{d\chi} d\chi \,,
\end{eqnarray}

\end{widetext}
where the last term is $\Omega_\phi$.
Fig.~\ref{fig:2} plots the average precession frequency for $\hat a=0.5$ and selected values of the eccentricity for stable bound orbits. The net average precession angle per azimuthal revolution for prograde orbits is then
\beq\label{deltadef}
 \Delta\Phi = \langle \Omega_{\rm prec}  \rangle \,T_\phi 
=   2\pi \frac{\langle \Omega_{\rm prec}  \rangle}{\Omega_\phi}  
\equiv   2\pi\left( 1-\delta \right)
\,.
\eeq


To second order in eccentricity we find

\begin{widetext}

\begin{eqnarray}\label{deltaorder2}
\delta 
 &=& \sqrt{1-3u_p+2\hat a u_p^{3/2}}\left[
1 -\frac32 \frac{ u_p^2 (1-2\hat a^3 u_p^{5/2}+3\hat a^2 u_p^2+2\hat a u_p^{3/2}-4u_p)(1-\hat a u_p^{1/2})^2}{(1-3 u_p+2\hat a u_p^{3/2})(1-6 u_p+8\hat a u_p^{3/2}-3\hat a^2 u_p^2) (\hat a^2 u_p^2-2 u_p+1)}e^2
\right]
+O(e^3)\nonumber\\
&\equiv&\delta^{(0)}+e^2\delta^{(2)}+O(e^3)\,.
\end{eqnarray}
The first term $\delta^{(0)}$ is the circular orbit limit which gives the number of revolutions in the angle $\Psi$ for one prograde revolution, so that 
\beq
\Delta\Phi^{(0)} = 2\pi(1-\delta^{(0)})
>0
\eeq
gives the net advance \cite{RindlerPerlick:1990,Iyer:1993qa}. 

Passing to the dimensionless (gauge-invariant) variable $y=(M\Omega_\phi)^{2/3}$ related to $u_p$ by 
\begin{eqnarray}
u_p &=& y'\left[
1+\left(
\frac23-\frac{1-3y'-2\hat a y'{}^{3/2}}{6(1-6y'+8\hat ay'{}^{3/2}-3\hat a^2y'{}^2)}
+\frac{1-5y'+4\hat a y'{}^{3/2}}{2(1-2y'+\hat a^2 y'{}^2)}
\right)e^2
\right]
+O(e^3)\,,
\end{eqnarray}
where $y'=y(1-\hat ay^{3/2})^{-2/3}$, we find (expanding in series of $y$ up to $O(y^7)$)
\begin{eqnarray}
\delta^{(0)} &=&  1-\frac32 y\left(1+\frac34 y+\frac{9}{8}y^2+\frac{135}{64}y^3+\frac{567}{128}y^4+\frac{5103}{512}y^5\right) \nonumber\\
&& +\hat a y^{3/2} \left(1+\frac12 y+\frac{15}{8}y^2+\frac{81}{16}y^3+\frac{1755}{128}y^4+\frac{9639}{256}y^5  \right) \nonumber\\
&& +\frac12 \hat a^2 y^3 \left(1-\frac76 y-\frac{37}{8}y^2-\frac{297}{16}y^3  \right) 
+O({\hat a}^3,e^3,y^7)\,,
\end{eqnarray}
and
\begin{eqnarray}
\delta^{(2)} &=& -\frac32 y \left(1+\frac12 y-\frac{1}{8}y^2-\frac{31}{16}y^3-\frac{901}{128}y^4-\frac{5305}{256}y^5 \right)  \nonumber\\
&&  +\frac32 \hat a  y^{3/2}\left(1-\frac76 y-\frac{103}{24}y^2-\frac{193}{16}y^3-\frac{11927}{384}y^4-\frac{59737}{768}y^5  \right) \nonumber\\
&& +\frac92 \hat a^2 y^3 \left( 1+\frac{46}{27}y+\frac{667}{216}y^2+\frac{41}{8}y^3\right)
+O({\hat a}^3,e^3,y^7) \,.
\end{eqnarray}

\end{widetext}

\section{Concluding remarks}

To the best of our knowledge past investigations of gyroscope precession have been limited to either geodesic or accelerated circular orbits in stationary axisymmetric spacetimes,
or general discussion using the language of relative observer analysis.
We have extended previous results to the case of gyroscopes moving along bound equatorial plane geodesic orbits in the Kerr spacetime. In this case it is natural and meaningful to consider an averaged precession frequency over a full (temporal) period of the motion.
We have shown how to identify natural axes in the local rest space of the gyro which are useful to measure the gyroscope precession and have shown how they differ only by a generalized Wigner rotation from those found \lq\lq by inspection" by Marck in his construction of a parallel propagated frame along a general geodesic in Kerr. We have explained the Lorentz geometry underlying the relationship between these two sets of axes, one tied to observations with respect to the distant stars, and the other to the Killing symmetries of the geodesic motion.
Moreover, we have provided exact expressions for both the precession frequency and its average over a radial period of the motion, as well as the the net rotation angle  after one such period. We expect that these expressions will be fundamental to any (forthcoming) generalization of the present  results for application to gravitational self-force corrections, namely when the back-reaction of the particle on the background can no longer be neglected.

\appendix

\section{Marck frame geometry}

The Marck frame found by \lq\lq inspection" for general values of $\theta$ can be explained by some orthogonal projection geometry together with the Carter frame alignment of the electric and magnetic parts of the
the Killing-Yano tensor 2-form.
The Carter family of fiducial observers have $4$-velocity $u_{({\rm car} )}$
and orthogonal spatial unit vector $E(\ucar)_{\hat \phi}$ in the Killing $2$-plane of $t$ and $ \phi$ given by
\begin{eqnarray}
u_{({\rm car} )}&=& \frac{r^2+a^2}{\sqrt{\Delta \Sigma}}\left(\partial_t +\frac{a}{r^2+a^2}\,\partial_\phi\right)
\,,\nonumber\\
u_{({\rm car} )}^\flat&=& \sqrt{\frac{\Delta}{\Sigma}}\left[-d t +a\sin^2\theta \,d\phi\right]
\,,\nonumber\\
E(\ucar)_{\hat \phi}&=&\frac{a\sin\theta}{\sqrt{\Sigma}}\left(\partial_t +\frac{1}{a\sin^2\theta}\,\partial_\phi\right) 
\,,\nonumber\\
E(\ucar)_{\hat \phi}^\flat&=&\frac{\sin\theta}{\sqrt{\Sigma}}\left[-a\,d t +(r^2+a^2)\,d\phi\right]\,,
\end{eqnarray}
respectively.
Note that $u_{({\rm car} )}$ and $E(\ucar)_{\hat \phi}$ are respectively future-oriented and rotating with positive angular velocity assuming $a>0$.
The vectors 
\beq
\label{carterframe}
E(\ucar)_{\hat r} =e_{\hat r}=\frac1{\sqrt{g_{rr}}}\,\partial_r\,,\quad
E(\ucar)_{\hat \theta} =e_{\hat \theta}=\frac1{\sqrt{g_{\theta \theta }}}\,\partial_\theta\,,
\eeq
together with $E(\ucar)_{\hat \phi}$ form an orthonormal spatial triad with dual 
\beq
W^\hr =\sqrt{g_{rr}}\,d r\,, \quad 
W^\htheta = \sqrt{g_{\theta \theta }} \,d \theta\,, \quad
W^\hphi =E(\ucar)_{\hat \phi}^\flat\,,
\eeq
adapted to Carter observers $u_{({\rm car} )}=E_0$ (with dual $W^0=-u_{({\rm car} )}^\flat$).

The unit tangent vector $U$ to the timelike geodesics \eqref{geosgen} has the following covariant and contravariant forms with respect to the Carter frame
\begin{eqnarray}
U^\flat&=& 
-E d t+\epsilon_r\frac{\sqrt{R(r)}}{\Delta} \,d r
+\epsilon_\theta\sqrt{\Theta(\theta)} \,d \theta+L \,d\phi
\,,\nonumber\\
U&=&
\frac{P}{\sqrt{\Delta \Sigma}} u_{({\rm car} )}
+\frac{\epsilon_r\sqrt{R(r)}}{\sqrt{\Delta \Sigma}}E(\ucar)_{\hat r}\nonumber\\
&&
+\frac{\epsilon_\theta\sqrt{\Theta(\theta)}}{\sqrt{\Sigma}}E(\ucar)_{\hat \theta}
+\frac{B}{\sin\theta\sqrt{\Sigma}} E(\ucar)_{\hat \phi}\,. \nonumber\\
\end{eqnarray}
The covariant components are separable functions of the Boyer-Lindquist coordinates. Introducing the Carter relative velocity and associated gamma factor of $U$
\begin{eqnarray}
U 
&=& \gamma\car [u_{({\rm car} )} +\nu\car^a E(\ucar)_a]\,,
\end{eqnarray}
we then have explicitly
\begin{eqnarray}
&&\gamma\car= \frac{P}{\sqrt{\Delta \Sigma}}\,,  \nonumber\\
&&\nu\car^a \,E(\ucar)_a=
\frac{\sqrt{\Delta }}{P}\left[  
\frac{\epsilon_r\sqrt{R(r)}}{\sqrt{\Delta }} \,E(\ucar)_{\hat r}\right.\nonumber\\
&&\left.\qquad
+ \epsilon_\theta\sqrt{\Theta(\theta)} \,E(\ucar)_{\hat \theta}
+\frac{B}{\sin\theta } \,E(\ucar)_{\hat \phi}\right]
\,.
\end{eqnarray}
Now further decompose this Carter relative velocity vector parallel and perpendicular to the Carter radial direction \cite{Bini:2011zzc} 
\begin{eqnarray}
\label{tree-vel-comp}
\nu\car^\Vert&=&\nu\car^\hr E(\ucar)_{\hat r}
\,, \nonumber\\
\nu\car^\perp &=&\nu\car^\htheta E(\ucar)_{\hat \theta} 
+\nu\car^\hphi E(\ucar)_{\hat \phi} \nonumber\\
&=&||\nu\car^\perp||\, \hat \nu\car^\perp
\,,
\end{eqnarray}
and the useful cross product quantity ($90$ degree rotation of $\nu\car^\perp$)
\begin{eqnarray}
\nu\car^\times &=&E(\ucar)_{\hat r}\times_{u_{\rm(car)}}\nu\car\nonumber\\
&=&\nu\car^\hphi E(\ucar)_{\hat \theta} -\nu\car^\htheta E(\ucar)_{\hat \phi}\nonumber\\ 
&=&||\nu\car^\times||\, \hat \nu\car^\times\,,
\end{eqnarray}
which is orthogonal to $\nu\car$ (and hence $U$) in the local rest space of $u_{\rm(car)}$.

Consider the Killing-Yano tensor with its electromagnetic-like decomposition in the Carter frame
\begin{eqnarray}
f&=&a\cos \theta \,\, [u_{({\rm car} )}^\flat \wedge W^\hr]+r\,\, [W^{\htheta\hphi}]\nonumber\\
&=& u_{({\rm car} )}^\flat \wedge {\mathcal E}(u_{({\rm car} )})+^*[u_{({\rm car} )}^\flat \wedge {\mathcal B}(u_{({\rm car} )})]\,.
\end{eqnarray}
using the notation $W^{\alpha\beta}=W^\alpha \wedge W^\beta$.
Then $f$ and its dual $f^*$ can be written as follows
\beq
f=-a\cos \theta \,\, W^{0\hr}+r\,\, W^{\htheta\hphi}\,,\qquad 
f^*= a\cos \theta \,\, W^{\htheta\hphi}+r\,\, W^{0\hr}\,.
\eeq
Carter's frame is very special because of it aligns both the parallel electric and magnetic fields ${\mathcal E}(u_{({\rm car} )})$ and ${\mathcal B}(u_{({\rm car} )})$  with the radial direction
\begin{eqnarray}
{\mathcal E}(u_{({\rm car} )})&=&a\cos \theta E(\ucar)_{\hat r}\equiv {\mathcal E} E(\ucar)_{\hat r}\,, \nonumber\\
{\mathcal B}(u_{({\rm car} )})&=&r E(\ucar)_{\hat r}\equiv {\mathcal B}E(\ucar)_{\hat r}\,,
\end{eqnarray}
having introduced the more compact notation
\beq
||{\mathcal E}(u_{({\rm car} )})||
=a|\cos \theta  |\equiv |{\mathcal E}|\,, \qquad 
||{\mathcal B}(u_{({\rm car} )})||=r \equiv {\mathcal B} \,.
\eeq
The invariants of this field
$
I_1= \frac12\,{\rm Tr}\, [f^2] = {\mathcal B}^2-{\mathcal E}^2$, $ I_2=\frac12\,{\rm Tr}\, [ff^*]  =2 {\mathcal E}{\mathcal B}
$
are both nonzero, showing that the field is nonsingular.

Marck \cite{marck1,marck2} takes the electric part of the Killing-Yano tensor with respect to the geodesic 4-velocity as a spacelike vector orthogonal to $U$ and parallely transported along $U$, given by
\begin{eqnarray}\label{e2marck}
e_2 &\propto&  
E(U) = f \rightcontract U \nonumber\\
 & =& {\mathcal E}( \nu\car^\hr u_{({\rm car} )} + E(\ucar)_{\hat r})+ {\mathcal B} \nu\car^\times
\,.
\end{eqnarray}
This only needs to be normalized to a unit vector.
Both of these two parts of $e_2$ are orthogonal to each other and to $U$, so by taking their corresponding unit vectors
\beq
\hat P= \gamma_\rad \left[\nu\car^\hr u_{({\rm car} )} + E(\ucar)_{\hat r} \right]\,,\qquad
\hat Q = \hat\nu\car^\times
\,,
\eeq
where $\gamma_\rad= (1-(\nu\car^\hr)^2)^{-1/2}$ is the ``partial" gamma factor associated with the radial motion alone needed to boost the radial direction into the local rest space of the gyro.

Let us introduce the following spherical component representation of the velocity components $\nu^a\car$
\begin{eqnarray}
\nu\car^\hr     &=&\nu\car \cos\alpha\,,\quad  
\nu\car^\htheta =\nu\car \sin\alpha \cos\beta\,,\nonumber\\  
\nu\car^\hphi   &=&\nu\car \sin\alpha \sin\beta \,.
\end{eqnarray}
We have then
\begin{eqnarray}
\gamma_\rad&=&(1-(\nu\car)^2 \cos^2\alpha)^{-1/2}\,,\nonumber\\
||\nu\car^\times||^2&=&\nu\car^2 \sin^2\alpha
\end{eqnarray}
and
\beq
[\sin\Theta, \cos \Theta] =  \frac{[{\mathcal E},{\mathcal B} \gamma_\rad\nu\car \sin \alpha ]}
{\sqrt{  {\mathcal E}^2+  {\mathcal B}^2 \gamma_\rad^2\nu\car^2 \sin^2\alpha }}
\,.
\eeq

The final form of $e_2$ is
\beq
   e_2 = \sin\Theta\, \hat P +\cos\Theta\, \hat Q\,,
\eeq
with 
\beq
\sin\Theta = \left({{\mathcal E}}/{ \gamma_\rad}\right)
 \left[\left({{\mathcal E}}/{ \gamma_\rad}\right)^2+  {\mathcal B}^2 ||\nu\car^\times||^2\right]^{-1/2}
\,,
\eeq
from which one easily obtains the third vector in this procedure up to a choice of sign
\beq
  \tilde e_1 =   \cos\Theta\, \hat P -\sin\Theta\, \hat Q\,,
\eeq
which is orthogonal to the previous one and to $U$ since both $\hat P$ and $\hat Q$ are orthogonal to $U$. The last vector of the Marck frame follows from orthogonality to $\tilde e_1$ and $e_2$ in the local rest space of $U$, namely $\tilde e_3 = \tilde e_1 \times_U e_2=\hat P\times_U\hat Q$. This discussion holds in general off the equatorial plane, completing the explanation of Ref.~\cite{Bini:2011zzc}.  

The relation \eqref{e2marck} of Marck has a simple interpretation as the transformation law for an electric field given as Eq.~(4.14)a in \cite{Jantzen:1992rg} with $(u,U)\to(U,\ucar)$
\begin{eqnarray}
 E(U)&=&\gamma(U,\ucar) P(\ucar,U)^{-1}(E(\ucar)\nonumber\\
  &&\quad + \nu(U,\ucar)\times_{\ucar}  {\mathcal B}(\ucar)) \,,
\end{eqnarray}
where $ P(\ucar,U)^{-1}$ is the inverse of the projection from the local rest space of $\ucar$ to that of $U$. Since these electric and magnetic vectors are parallel and radial in the Carter frame, this implies
\begin{eqnarray}
 E(U)&=&\gamma(U,\ucar) P(\ucar,U)^{-1}
  \left[{\mathcal E} E(\ucar)_\hr \right.\nonumber\\
&& \ 
+ \left.{\mathcal B} \nu(U,\ucar)\times_{\ucar}  E(\ucar)_\hr \right]\,.
\end{eqnarray}
The first term using Eq.~(4.7) of \cite{Jantzen:1992rg} evaluates to
\beq
P(\ucar,U)^{-1}
 E(\ucar)_\hr = 1/\gamma_\rad E(u_\rad)_\hr
\eeq
while the second term is unchanged by the projection, giving finally
\begin{eqnarray}
 E(U)&=&\gamma(U,\ucar) 
  \left[\frac{{\mathcal E}}{\gamma_\rad} E(u_\rad)_\hr \right.\nonumber\\
&&\left.
+ {\mathcal B} \nu(U,\ucar)\times_{\ucar}  E(\ucar)_\hr \right]\,.
\end{eqnarray}
These two terms are orthogonal and so define the two unit vectors $\hat P$ and $\hat Q$ respectively, and normalizing their sum defines $e_2$ and the angle $\Theta$ needed to get $\tilde e_1$. Thus the Lorentz geometry of the Killing-Yano form underlies this previously unexplained Marck procedure.

We now specialize the discussion to equatorial plane orbits
where $\mathcal E=0=\nu\car^\htheta$, so $\Theta=0$ and
$e_2$ is aligned with the $\theta$ direction while  $\tilde e_1$ is aligned with a boost of the radial direction into the local rest space of the geodesic, leaving $\tilde e_3$ along the boosted azimuthal direction. Explicitly
\beq
\tilde e_1 = \hat P\,,\quad
e_2 = \hat Q\,,\quad
\tilde e_3 = \hat P \times_U\hat Q
\,.
\eeq
These two vectors can be understood as the result of three successive relative observer boosts from the local rest space of the static observer $m$ \cite{Jantzen:1992rg}. The first is an azimuthal boost $B(u\car,m)$ from $m$ to $u\car$, the second is $B(u\carrad,u\car)$ from Carter along the radial direction, followed by the third $B(U,u\carrad)$ to the local rest space of the gyro, where
\begin{eqnarray}
  u\carrad&=& \gamma\carrad (u\car+ \nu\car^{\hat r}  E(u\car)_{\hat r}) \,,\nonumber\\
 E(m)_{\hat r} &=& E(u\car)_{\hat r} 
\end{eqnarray}
is the result of boosting the Carter observer in the radial direction to comove radially with the gyro, leaving the azimuthal direction invariant. 
In this intermediate frame the gyro relative velocity only has an angular component, which in the equatorial plane case reduces to  the azimuthal frame component
\begin{eqnarray}
 && U = \gamma(U,u\carrad) \left[ u\carrad + \nu(U,u\carrad)^{\hat \phi} E(u\carrad)_{\hat \phi}  \right] \,,\nonumber\\
 && E(u\carrad)_{\hat \phi} = E(u\car)_{\hat \phi}\,,
\end{eqnarray}
where $ \gamma(U,u\carrad)=  \gamma\car/\gamma_\rad$
and the final boost $B(U,u\carrad)$ leaves the radial direction invariant.
This sequence of boosts is

\begin{widetext}

\beq\label{boosts}
\left(\begin{array}{c} \tilde e_1 \\ \tilde e_3 \end{array}\right)
= B(U,u\carrad)\, B(u\carrad,u\car)\, B(u\car,m)\,
\left(\begin{array}{c} E(m)_{\hat r} \\ E(m)_{\hat \phi} \end{array}\right)
\,.
\eeq

\end{widetext}
Note that radial boost $B(u\carrad,u\car)$ leaves invariant the unit area 2-form 
\beq
  u\carrad \wedge \tilde e_1 = u\car \wedge E(m)_{\hat r}
\eeq
in the $u\car$-$E(u\car)_{\hat r}$ subspace of the tangent space, as well as the orthogonal 2-form, thus leaving the electric and magnetic 2-form parts of the Killing-Yano 2-form invariant.

We need to compare these axes to the direct boost $B(U,m)$ from the static observers to the geodesic 
\beq\label{boosts}
\left(\begin{array}{c} E(U)_{\hat r} \\ E(U)_{\hat \phi} \end{array}\right)
= B(U,m)\,
\left(\begin{array}{c} E(m)_{\hat r} \\ E(m)_{\hat \phi} \end{array}\right)
\,.
\eeq
Note that the overall boost $B(U,m)$ has the effect of removing the stellar aberration of the incoming light rays from the \lq\lq fixed stars" at radial infinity whose unit relative velocities (direction vectors) are aligned with the static observer local rest space directions.

The three successive boosts lead to a generalized Wigner rotation compared to the direct boost, a kinematical effect which only depends on the relative velocity with respect to the static observers much like the stellar aberration effect which instead is due to the projection of a unit relative velocity between two local rest frames

\begin{widetext}

\beq\label{Buurad}
 B(U,u\carrad)\, B(u\carrad,u\car)\, B(u\car,m) = R_{\rm(wig)}(U,u\carrad,u\car,m) \, B(U,m)\,,
\eeq
where $R_{\rm(wig)}(U,u\carrad,u\car,m)$ is the product of two ordinary Wigner rotations
\beq
  R_{\rm(wig)}(U,u\carrad,u\car,m) = R_{\rm(wig)}(U,u\carrad,u\car) \, R_{\rm(wig)}(U,\ucar,m) \,.
\eeq

\end{widetext}

This same boost discussion applies to the general case of nonequatorial plane motion as well, taking into account the full angular relative velocity, thus explaining the geometric origin of Marck's choice of frame in that context as well.
We calculate the generalized Wigner rotation in Appendix C for equatorial plane motion, but it does not contribute to the average precession per radial period. Curiously this interesting geometry has never been explored before. The intermediate relative observer $u\carrad$ is the key to this calculation, coming from Marck's derivation making use of the Killing-Yano 2-form, but the rest is straightforward though nontrivial Lorentz geometry.

Note that one could have reversed the order of the radial and angular boosts from the Carter frame to boost first in the angular direction from the Carter observer to 
\begin{eqnarray}\label{car-ang}
  u_{\rm(ang)} &=& \gamma_{\rm(ang)} \left[\ucar+\nu\car^\hphi E(\ucar)_\hphi\right] \,,
\nonumber\\
  E(u_{\rm(ang)})_\hphi &=& \gamma_{\rm(ang)} \left[\nu\car^\hphi \ucar+E(\ucar)_\hphi\right] \,,\nonumber\\
\end{eqnarray}
where  $  \gamma_{\rm(ang)} =(1-|| \nu\car^\hphi||^2)^{-1/2}$.
The new angular vector in this transition has a formula analogous to \eqref{marckframe13} with \eqref{car-m-r} backsubstituted.
This boost combines additively with the boost from the static observer to the Carter observer since they are both in the same plane 
\beq
B(u\carang,u\car)\, B(u\car,m) = B(u\carang,m)\,,
\eeq
which is
then followed by the final boost in the radial direction to $U$. The resulting rotation relative to the direct boost would be a true Wigner rotation
\beq\label{Buuang}
 B(U,u\carang)\, B(u\carang,m) = R_{\rm(wig)}(U,u\carang,m) \, B(U,m)\,.
\eeq
However, the construction starting first with the radial boost is preferred because of the common radial direction of the electric and magnetic vector fields associated with the Killing-Yano 2-form.

\section{Transformation law for the cross product between different local rest spaces}

The final vector in the Marck frame must be calculated with the local rest space cross product, for which a useful evaluation formula can be derived using the projection formalism of Refs.~\cite{Jantzen:1992rg,ferrbini}.
Let LRS$_U$ and LRS$_u$ the local rest spaces associated with the two unit timelike vector fields $U$ and $u$, related by the boost
\beq
U=\gamma(U,u)[u+\nu(U,u)]\,.
\eeq
The cross product 
\beq
[X\times_U Y]^\alpha = \eta(U)^{\alpha\beta\gamma} X_\beta Y_\gamma
\eeq
in LRS$_U$  is defined for generic vectors $X$ and $Y$ using 
$\eta(U)^{\alpha\beta\gamma}=U_\sigma \eta^{\sigma\alpha\beta\gamma}$ 
with similar defining relations in LRS$_u$. $\eta_{\sigma\alpha\beta\gamma}$ is the unit volume 4-form, whose components in an oriented orthonormal frame are fixed to be $\eta_{0123}=1$.
This leads to the relation
\begin{eqnarray}
\label{gen_rel}
[X\times_U Y]&=&\gamma(U,u)\{ [X\times_u Y]+u (\nu(U,u)\cdot [X\times_u Y])\nonumber\\
&&
+(X\cdot u)[\nu(U,u)\times_u Y]\nonumber\\
&&
-(Y\cdot u)[\nu(U,u)\times_u X]
\}\,,
\end{eqnarray}
which in the special case of $X,Y \in$ LRS$_u$ reduces to
\beq
[X\times_U Y]=\gamma(U,u)\left\{ [X\times_u Y]+u (\nu(U,u)\cdot [X\times_u Y])
\right\}\,.
\eeq

Recall the definition of the projector $P(U)$ orthogonal to the timelike unit vector $U$,
)($P(U)^\flat=g+U^\flat\otimes U^\flat$) we see that the vectors $X$ and $Y$ in the cross product in the local rest space of $U$, $\times_U$, can be equivalently replaced by
$P(U)X$ and $P(U)Y$; similarly, the vectors $X$ and $Y$ in the cross product in the local rest space of $u$, $\times_u$, can be equivalently replaced by
$P(u)X$ and $P(u)Y$.

To study an application of Eq.~\eqref{gen_rel} let us write $U=\gamma (u+\nu^a E(u)_a)$ with the abbreviations
$\gamma(U,u)\to\gamma$, $\nu(U,u)\to\nu$
and introduce also
the three vectors
\begin{eqnarray}
\nu^\Vert&=&\nu^1 E(u)_1\,,\nonumber\\ 
\nu^\perp &=& \nu^2 E(u)_2+\nu^3 E(u)_3=||\nu^\perp||\hat \nu^{\perp}\,,\nonumber\\
\nu^\times&=&\nu^3 E(u)_2-\nu^2 E(u)_3=||\nu^\times|| \hat \nu^\times\,,
\end{eqnarray}
in the local rest space of $u$.
Let us consider the two vectors in the local rest space of $U$
\beq
X=\nu^1 u+E(u)_1\,,\qquad Y=\nu^\times
\eeq
satisfying $u\cdot X=-\nu^1$ and $u \cdot Y=0$.
We have then
\begin{eqnarray}
X\times_U Y&=& \gamma ||\nu^\perp||\left\{  \left( ||\nu^\perp|| u +\hat \nu^\perp\right)\right.\nonumber\\
&&\left.
+\nu^1 \left( ||\nu^\perp|| E(u)_1 -\nu^1 \hat \nu^\perp\right)  \right\}\,,
\end{eqnarray}
with each of the two vectors $||\nu^\perp|| u +\hat \nu^\perp$ and  $||\nu^\perp|| E(u)_1 -\nu^1 \hat \nu^\perp$ orthogonal to  $U$. Then the direction formula is
\beq
\frac{X\times_U Y}{||X\times_U Y||}= \frac{\gamma_1}{\gamma} 
\left(||\nu^\perp||\gamma U+\hat \nu^\perp\right)=\frac{P(U)\hat \nu^\perp}{||P(U)\hat \nu^\perp||}\,,
\eeq
with $\gamma_1=(1-(\nu^1)^2)^{-1/2}$ and $||P(U)\hat \nu^\perp||= \gamma/\gamma_1$.

This formula will easily allow for the computation of terms like $\hat P \times_U \hat Q$, as introduced in the previous sections, where $u=u_{\rm(car)}$ and $U$ is the geodesic 4-velocity
\beq
\hat P\times_U \hat Q = \frac{\gamma_\rad}{\gamma\car} 
\left(||\nu\car^{\hat \phi}||\gamma\car U+ E(\ucar)^{\hat \phi}\right)\,.
\eeq

\section{Wigner rotation}

The Carter observers play a key role in the geodesic motion and parallel transport along those orbits, while the static observers are key to defining spin precession as seen from radial infinity. We compare the Marck frame vectors with the boosted static observer frame vectors to see the relative rotation between them for equatorial plane motion.

The geodesic 4-velocity can be decomposed into relative motion with respect to the static observers with 4-velocity $m$ and the Carter observers with 4-velocity $u_{({\rm car} )}$, whose distinct frame vectors are
\begin{eqnarray}
  m &=&  \frac{1}{N} \partial_t \,,\qquad
 E(m)_{\hat \phi} = -\frac{2aM}{r N\sqrt{\Delta }} \partial_t + \frac{N}{\sqrt{\Delta}} \partial_\phi\,.\nonumber\\
\end{eqnarray}
with $N=\sqrt{1-2M/r}$ and
\begin{eqnarray}
u_{({\rm car} )}&=& \frac{r^2+a^2}{r\sqrt{\Delta}}\left(\partial_t +\frac{a}{r^2+a^2}\partial_\phi\right)\,,\nonumber\\
E(u_{(\rm car)})_{\hat \phi}&=&\frac{a}{r}\left(\partial_t +\frac{1}{a}\partial_\phi\right)\,,
\end{eqnarray}
and whose common frame vectors are
\begin{eqnarray}
E(u_{(\rm car)})_{\hat r} &=& E(m)_{\hat r} \equiv e_{\hat r}= \frac{\sqrt{\Delta}}{r} \partial_r\,,\nonumber\\
E(u_{(\rm car)})_{\hat \theta} &=& E(m)_{\hat \theta}\equiv e_{\hat \theta} =\frac{1}{r} \partial_\theta\,,
\end{eqnarray}
leading to relative velocities and gamma factors such that
\begin{eqnarray}
U&=&\gamma \left[m +\nu^{\hat r}e_{\hat r}+\nu^{\hat \phi}E(m)_{\hat \phi}\right]\nonumber\\
&=&\gamma\car  \left[u_{(\rm car)} +\nu\car ^{\hat r}e_{\hat r}+\nu\car ^{\hat \phi}E(u_{(\rm car)})_{\hat \phi}\right] \,,
\end{eqnarray}
with equal radial components
\beq\label{car-m-r}
\gamma \nu^{\hat r} = \gamma\car  \nu\car ^{\hat r} 
    =\frac{r \dot r}{\sqrt{\Delta}} \,,
\eeq
and
\begin{eqnarray}
\gamma &=& \frac{E}{N} \,,\qquad\qquad
\gamma \nu^{\hat \phi} =\frac{L r -2M x}{rN \sqrt{\Delta}} \,,\nonumber\\
\gamma\car  &=& \frac{Er^2-ax}{r\sqrt{\Delta}} \,,\quad
 \gamma\car \nu\car ^{\hat \phi} = \frac{x}{r}\,.
\end{eqnarray}
In turn we can decompose the Carter 4-velocity
\beq
u_{(\rm car)} = \gamma_{c,m} \left[m+ \nu^{\hat \phi}_{c,m} E(m)_{\hat \phi} \right]
\eeq
with
\begin{eqnarray}
 \gamma_{c,m} &=& \frac{\sqrt{\Delta}}{rN} \,,\quad
\nu^{\hat \phi}_{c,m} = \frac{a}{\sqrt{\Delta}}\,.
\end{eqnarray}
and
\beq
 \gamma= \gamma\car \gamma_{c,m} \left( 1+ \nu\car^{\hat \phi} \nu_{c,m}^{\hat \phi} \right)\,,\qquad
 \nu^{\hat \phi} = \frac{ \nu_{c,m}^{\hat \phi} + \nu\car^{\hat \phi} }{1+\nu_{c,m}^{\hat \phi} \nu\car^{\hat \phi}} \,.
\eeq

We recall the notation for the general relative observer boost map between two different local rest spaces \cite{Jantzen:1992rg}, one orthogonal to $U$ and the other to $u$, with $U=\gamma(U,u)[u+\nu(U,u)]$ and the reciprocal relation $u=\gamma(u,U)[U+\nu(u,U)]$ with $\gamma(U,u)=\gamma(u,U)\equiv \gamma$.
For a vector $X\in LRS_u$, the vector boosted into $LRS_U$ in the plane of $u$ and $U$ is given by the right contraction with the projection $P(U)$ from $LRS_u$ to $LRS_U$ acting on $X$

\begin{widetext}

\begin{eqnarray}
\label{general_boost}
B(U,u)X  &=& \left(P(U) +\frac{\gamma}{\gamma+1} \nu(u,U)\otimes \nu(u,U)^\flat\right)\rightcontract (P(U)X)\nonumber\\
&=& X+\frac{\gamma}{\gamma+1}(\nu(U,u)\cdot X) (u+U)\,.
\end{eqnarray}
For example, by using Eq.~\eqref{general_boost} it is easy to show that
\beq
B(u_{(\rm car)},m)E(m)_{\hat \phi}=E(u_{(\rm car)})_{\hat \phi}\,.
\eeq
In fact
\begin{eqnarray}
B(u_{(\rm car)},m)E(m)_{\hat \phi}&=& E(m)_{\hat \phi}+\frac{\gamma_{c,m}}{\gamma_{c,m}+1}\nu^{\hat \phi}_{c,m} (u_{(\rm car)}+m)\nonumber\\
&=& \gamma_{c,m} \nu^{\hat \phi}_{c,m} m +\left(1+\frac{\gamma_{c,m}^2 (\nu^{\hat \phi}_{c,m})^2}{\gamma_{c,m}+1}  \right) E(m)_{\hat \phi}\nonumber\\
&=& \gamma_{c,m} [\nu^{\hat \phi}_{c,m} m + E(m)_{\hat \phi} ]\nonumber\\
&=& \frac{a}{r}\left(\partial_t +\frac{1}{a}\partial_\phi\right)\,.
\end{eqnarray}

\end{widetext}
Next we consider the static observer frame vectors boosted to the local rest space of $U$
\begin{eqnarray}
E(U)_{\hat r} &=& B(U,m) E(m)_{\hat r} = e_{\hat r} + \frac{\gamma\, \nu^{\hat r} }{\gamma+1} (m+U) \,,\nonumber\\
E(U)_{\hat \phi} &=& B(U,m) E(m)_{\hat \phi} = E(m)_{\hat \phi} + \frac{\gamma\, \nu^{\hat \phi} }{\gamma+1} (m+U) \,,\nonumber\\
\end{eqnarray}
which must be compared to the Marck frame vectors
\begin{eqnarray}\label{marckframe13}
\tilde e_1&=&\hat P = \gamma_\rad \left( \frac{\gamma\, \nu^{\hat r}}{\gamma\car } u_{(\rm car)} +e_{\hat r} \right)  \,,\nonumber\\
\tilde e_3 &=&\hat P \times_U \hat Q = \frac{\gamma_\rad}{\gamma\car} 
\left(\nu^{\hat \phi}\car\gamma\car U+E(u\car)_{\hat\phi}\right) \,,\nonumber\\
\end{eqnarray}
where $\hat Q=e_{\hat \theta}$.

The boosted axes are rotated with respect to the Marck axes by a counterclockwise rotation by an angle $\Lambda$

\begin{widetext}

\beq\label{rotlamba}
\left(\begin{array}{c} E(U)_{\hat r} \\ E(U)_{\hat \phi} \end{array}\right)
= R(-\Lambda) \left(\begin{array}{c} \tilde e_1 \\ \tilde e_3 \end{array}\right)
=\left(\begin{array}{cc}  \cos \Lambda & \sin \Lambda \\ -\sin \Lambda & \cos \Lambda\end{array}\right)
\left(\begin{array}{c} \tilde e_1 \\ \tilde e_3 \end{array}\right)
\,.
\eeq
One finds
\begin{eqnarray}
  \cos\Lambda &=& \hat P \cdot E(U)_{\hat r} 
  = \gamma_\rad \left( 1-\frac{\gamma_{c,m}}{\gamma_{(\rm car)}} \frac{\gamma^2 (\nu^{\hat r})^2}{\gamma+1}\right)\nonumber\\
&=& \left(1-\frac{r^4 \dot r^2}{(E r^2 -ax)^2}  \right)^{-1/2}  \left[1-  \frac{r^2\dot r^2}{(E r^2+ax)(E-N)}\right]\nonumber\\
&=& \frac{1}{\sqrt{\Delta (r^2+x^2)}}\left[r^2N+ax-\frac{(a-Nx^2)^2}{E+N}  \right]\,.
\end{eqnarray}
Similarly
\begin{eqnarray}
 \sin\Lambda &=&
(\hat P\times_U \hat Q) \cdot E(U)_{\hat r} 
= \left\{P(U)[(\hat P\times_U \hat Q)]\right\} \cdot E(U)_{\hat r}
=
\frac{\gamma_\rad}{\gamma\car} E(u\car)_{\hat\phi} \cdot E(U)_{\hat r}\nonumber\\
&=& \nu^{\hat r}\,\frac{\gamma\gamma_\rad \gamma_{c,m}}{\gamma\car}  \left( \frac{\gamma \nu^{\hat \phi}}{\gamma+1}-\nu_{c,m}^{\hat \phi} \right)
 \,.
\end{eqnarray}

\end{widetext}
Note that at either periastron or apastron the radial relative velocity vanishes so $\Lambda=0$, aligning the two sets of orthonormal vectors there.

Then we have the sequence of rotations and boost
\beq
\left(\begin{array}{c} e_1 \\ e_3 \end{array}\right)
= R(\Psi) \, R(\Lambda) B(U,m)
\left(\begin{array}{c} e_{\hat r} \\ E(m)_{\hat \phi} \end{array}\right)
\,,
\eeq
If we introduce the Cartesian-like axes which are rotated clockwise by the angle $\phi$ relative to $\phi=0$ along the orbit, to remove the counterclockwise rotation associated with the increasing variable $\phi$
\beq
\left(\begin{array}{c} E_{\hat x} \\ E_{\hat y}\end{array}\right)
= R(\phi) 
\left(\begin{array}{c} e_{\hat r} \\ E(m)_{\hat \phi} \end{array}\right)
\,,
\eeq
we get finally
\begin{eqnarray}
\left(\begin{array}{c} e_1 \\ e_3 \end{array}\right)
&=& R(\Psi) \, R(\Lambda) B(U,m) R(-\phi) 
\left(\begin{array}{c} E_{\hat x} \\ E_{\hat y} \end{array}\right)
\nonumber\\
&=& R(\Psi-\phi+\Lambda) \, B(U,m)  
\left(\begin{array}{c} E_{\hat x} \\ E_{\hat y} \end{array}\right)
\,,
\end{eqnarray}
since the rotation of the orthonormal pair of vectors in the equatorial plane commutes with the boost which fixes the radial direction. Thus one only need to add the term $d\Lambda/dt$ to the precession formula \eqref{omegaprec} to get the instantaneous precession formula but this does not contribute to the average over one radial period.

Since the angle $\Lambda$ of the Marck axes is just a function of the relative velocity which is odd in the radial component, it is starts at zero at the periastron and is positive during the half orbit from periastron to aphelion returning to zero at the aphelion, and then is negative on the return to the periastron where it again returns to zero (so that at the extreme radii the axes are aligned with the spherical axes). This is similar to stellar aberration in some sense, where the direction of a fixed star around an orbit has a similar periodic oscillation with respect to the simple rotation.
Indeed one can verify that
\beq
  \int_0^{2\pi} \frac{d\Lambda}{d\tau} \frac{d\tau}{d\chi} \, d\chi =0\,.
\eeq

\subsection*{Acknowledgments}
D.B. thanks the Italian INFN (Naples) for partial support and Prof.~B. Mashhoon at the University of Columbia Missouri (USA) for informative discussion on the construction of a parallely propagated frame along general geodesics in the Kerr spacetime.
All  the authors are grateful to the International Center for Relativistic Astrophysics Network based in Pescara, Italy for partial support.

\end{document}